\documentclass[sigconf]{acmart}


\usepackage{booktabs} 
\usepackage{graphicx,url}
\usepackage{subfigure}
\usepackage{booktabs}
\usepackage{balance}
\usepackage{array} 
\usepackage{varwidth} 
\usepackage{scalefnt}
\usepackage{marginnote}
\usepackage{tcolorbox}

\newcommand{\aspas}[1]{{``#1''}}

\definecolor{gray}{RGB}{90,90,90}
\newcommand\sbar[2]{{\color{gray}\rule{\dimexpr 1cm * #1 / #2}{6pt}}}

\newcommand{\lenghtReasons}[0]{0.48}
\newcommand{\lenghtForks}[0]{0.22}

\newcommand{\totalGitHubUsers}[0]{19 million}
\newcommand{\totalGitHubRepositories}[0]{52 million}

\newcommand{\totalStudiedProjects}[0]{618}
\newcommand{\totalProjectsWithoutCommitsBeforeFilter}[0]{628}
\newcommand{\totalProjectsWithoutCommitsBeforeFilterRate}[0]{13}
\newcommand{\totalLibrariesAndFrameworkStudiedProject}[0]{502}
\newcommand{\totalLibrariesAndFrameworkStudiedProjectRate}[0]{81}
\newcommand{\totalStudiedProjectsWithoutCommits}[0]{542}
\newcommand{\totalStudiedProjectsWithDeprecatedMessages}[0]{76}
\newcommand{\totalPersonalProjects}[0]{448}
\newcommand{\totalPersonalProjectsRate}[0]{72}
\newcommand{\totalOrganizationProjects}[0]{170}
\newcommand{\totalOrganizationProjectsRate}[0]{28}

\newcommand{\medianMonthsAllProjects}[0]{40}
\newcommand{\medianMonthsStudiedProjects}[0]{52}
\newcommand{\medianContributorsAllProjects}[0]{23}
\newcommand{\medianContributorsStudiedProjects}[0]{11}
\newcommand{\medianCommitsAllProjects}[0]{346}
\newcommand{\medianCommitsStudiedProjects}[0]{137}
\newcommand{\medianStarsAllProjects}[0]{2,538}
\newcommand{\medianStarsStudiedProjects}[0]{2,345}

\newcommand{\removedIsNotSoftwareProject}[0]{51}
\newcommand{\removedNativeLanguageIsNotEnglish}[0]{24}
\newcommand{\removedMovedToAnotherRepository}[0]{7}
\newcommand{\removedRepositoryEmpty}[0]{4}

\newcommand{\developersSurveyedByEmail}[0]{118}

\newcommand{\developersSurveyedByReadme}[0]{36}

\newcommand{\developersWithValidAndPublicEmail}[0]{425}
\newcommand{\developersOwnerOrContributorOfTwoOrMoreProjects}[0]{9}
\newcommand{\developersKeepsMaintenanceOutOfRepository}[0]{two}
\newcommand{\developersAgreed}[0]{101}
\newcommand{\developersAgreedRate}[0]{86}
\newcommand{\developersNotAgreed}[0]{17}
\newcommand{\developersNotAgreedRate}[0]{14}
\newcommand{\totalDevelopersSurveyed}[0]{154}
\newcommand{\totalSentEmails}[0]{414}
\newcommand{\totalEmailsReturnedWithProblem}[0]{6}
\newcommand{\responseRateSurvey}[0]{29}

\newcommand{\totalReasonOfLackOfTime}[0]{27}
\newcommand{\totalReasonOfLackOfInterest}[0]{30}
\newcommand{\totalReasonOfUsurpedByCompetitor}[0]{30}
\newcommand{\totalReasonOfCompleted}[0]{17}
\newcommand{\totalReasonOfOutdatedTehcnologies}[0]{16}
\newcommand{\totalReasonOfObsolete}[0]{21}
\newcommand{\totalReasonOfTechnicalReason}[0]{7}
\newcommand{\totalReasonOfUnclear}[0]{five}
\newcommand{\totalReasonOfConflictsAmongDevelopers}[0]{three}
\newcommand{\totalReasonOfAcquisition}[0]{two}
\newcommand{\totalReasonOfLackOfPatience}[0]{one}
\newcommand{\totalReasonOfLackOfExpertise}[0]{one}
\newcommand{\totalReasonOfLegalProblems}[0]{two}

\newcommand{\totalReasonOfCompletedRate}[0]{11}

\newcommand{\totalReasonWhyOSSProjectFail}[0]{nine}

\newcommand{\totalReasonWhyFailOfLackOfTime}[0]{18}
\newcommand{\totalReasonWhyFailOfLackOfInterest}[0]{18}
\newcommand{\totalReasonWhyFailOfUsurpedByCompetitor}[0]{27}
\newcommand{\totalReasonWhyFailOfOutdatedTehcnologies}[0]{14}
\newcommand{\totalReasonWhyFailOfObsolete}[0]{20}
\newcommand{\totalReasonWhyFailOfTechnicalReason}[0]{7}
\newcommand{\totalReasonWhyFailOfConflictsAmongDevelopers}[0]{3}
\newcommand{\totalReasonWhyFailOfAcquisition}[0]{1}

\newcommand{\totalReasonWhyFailOfLegalProblems}[0]{2}

\newcommand{\totalReasonDevelopmentTeam}[0]{39}
\newcommand{\totalReasonProjectCharacteristics}[0]{41}
\newcommand{\totalReasonExternalEnvironment}[0]{30}
\newcommand{\totalReasonFailsSum}[0]{110}

\newcommand{\totalFundingCompanyEmploying}[0]{12}
\newcommand{\totalFundingPrivateCompany}[0]{two}
\newcommand{\totalFundingNonProfitOrganizations}[0]{three}

\newcommand{\totalProjectsFailuresReasonsAttendQuestionOne}[0]{76}
\newcommand{\totalProjectsFailuresReasonsAttendQuestionTwo}[0]{32}
\newcommand{\totalRemovedProjectsByHaveFewCommits}[0]{four}
\newcommand{\totalProjectsFails}[0]{104}

\newcommand{\totalRubyProjectsFails}[0]{10}
\newcommand{\totalRubyProjectsFailsWithoutDependents}[0]{6}
\newcommand{\totalJavaScriptProjectsFails}[0]{28}
\newcommand{\totalJavaScriptRubyProjectsFails}[0]{38}
\newcommand{\totalJavaScriptProjectsFailsOutNPM}[0]{10}
\newcommand{\totalJavaScriptProjectsWithFiveOrLessDependents}[0]{15}
\newcommand{\totalJavaScriptProjectsWithFiveOrLessDependentsRate}[0]{55}

\newcommand{\totalDependentsTopOneJavaScriptProjectsFails}[0]{158}
\newcommand{\totalDependentsTopTwoJavaScriptProjectsFails}[0]{37}
\newcommand{\totalDependentsTopThreeJavaScriptProjectsFails}[0]{13}

\newcommand{\totalDependentsTopOneRubyProjectsFails}[0]{2,460}
\newcommand{\totalDependentsTopTwoRubyProjectsFails}[0]{270}
\newcommand{\totalDependentsTopThreeRubyProjectsFails}[0]{36}
\newcommand{\totalDependentsTopFourRubyProjectsFails}[0]{18}

\newcommand{\totalProjectsFailWithMultipleReasons}[0]{6}

\newcommand{\totalProjectsNotFunding}[0]{82}
\newcommand{\totalProjectsNotFundingRate}[0]{69}

\newcommand{\developersHavePlansToReactivateTheProject}[0]{18}
\newcommand{\developersHavePlansToReactivateTheProjectRate}[0]{15}

\newcommand{\totalUsurpedByGoogle}[0]{7}
\newcommand{\totalUsurpedByApple}[0]{5}

\newcommand{\medianIssuesOpenedAll}[0]{18}
\newcommand{\medianIssuesOpenedDevelopmentTeam}[0]{43}
\newcommand{\medianIssuesOpenedProjectCharacteristics}[0]{21}
\newcommand{\medianIssuesOpenedExternalEnvironment}[0]{12}

\newcommand{\medianPullsOpenedAll}[0]{5}
\newcommand{\medianPullsOpenedDevelopmentTeam}[0]{11}
\newcommand{\medianPullsOpenedProjectCharacteristics}[0]{4}
\newcommand{\medianPullsOpenedExternalEnvironment}[0]{4}

\newcommand{\topOneIssuesOpened}[0]{230}
\newcommand{\topTwoIssuesOpened}[0]{173}
\newcommand{\topThreeIssuesOpened}[0]{160}

\newcommand{\topOnePullsOpened}[0]{54}
\newcommand{\topTwoPullsOpened}[0]{45}
\newcommand{\topThreePullsOpened}[0]{38}

\newcommand{\issuesTotalProjectsWichNotManageByGitHub}[0]{15}
\newcommand{\issuesTotalFailProjectsWichManageIssuesByGitHub}[0]{89} 
\newcommand{\issuesTotalIssuesAboutProjectStatus}[0]{32}

\newcommand{\issuesTotalStrategies}[0]{three}
\newcommand{\issuesSuggestsSubstitute}[0]{19}
\newcommand{\issuesSuggestsOrganization}[0]{five}
\newcommand{\issuesSuggestsNewMaintainer}[0]{three}
\newcommand{\issuesSuggestsAddCollaborators}[0]{five}
\newcommand{\issuesSuggestsNewMaintainerAndFound}[0]{two}

\newcommand{\forksFirstQuartile}[0]{244}
\newcommand{\forksMedian}[0]{400}
\newcommand{\forksThirdQuartile}[0]{638}

\newcommand{\forksThirdQuartileOfStarOfBestFork}[0]{13}

\setcopyright{acmcopyright}
\acmDOI{10.1145/3106237.3106246}
\acmISBN{978-1-4503-5105-8/17/09}
\acmConference[ESEC/FSE'17]{2017 11th Joint Meeting of the European Software Engineering Conference and the ACM SIGSOFT Symposium on the Foundations of Software Engineering}{September 4--8, 2017}{Paderborn, Germany} 
\acmYear{2017}
\copyrightyear{2017}
\acmPrice{15.00}

\begin{document}


\title{Why Modern Open Source Projects Fail}

\author{Jailton Coelho, Marco Tulio Valente}
\affiliation{%
  \institution{Federal University of Minas Gerais}
  \streetaddress{Department of Computer Science}
  \city{Belo Horizonte}
  \state{Minas Gerais, Brazil}
}
\email{{jailtoncoelho,mtov}@dcc.ufmg.br}


\begin{abstract}
Open source is experiencing a renaissance
period, due to the appearance of modern platforms and workflows for developing and
maintaining public code. As a result, developers are creating open source software at
speeds never seen before. Consequently, these projects are also facing unprecedented
mortality rates. To better understand the reasons for
the failure of modern open source projects, this paper describes the results
of a survey with the maintainers of \totalProjectsFails\  popular GitHub systems that have been
deprecated. We provide a set of \totalReasonWhyOSSProjectFail\ reasons for the failure of these open source projects.
We also show that some maintenance practices---specifically the adoption of
contributing guidelines and continuous integration---have an important
association with a project failure or success. Finally, we discuss and reveal the
principal strategies developers have tried to overcome the failure
of the studied projects.
\end{abstract}


\maketitle

\section{Introduction}

 Over the years, the open source movement is contributing to a dramatic
reduction in the costs of building and deploying software. Today, organizations often rely on
open source to support their basic software infrastructures, including operating systems,
databases, web servers, etc. Furthermore, most software produced nowadays depends on
public source code, which is used for example to encapsulate the implementation of code related to security,
 authentication, user interfaces, execution on mobile devices, etc.  
 A recent survey shows that 65\% out of 1,313 surveyed companies rely on open source to
 speed application development.\footnote{\url{https://www.blackducksoftware.com/2016-future-of-open-source}}
For example,  Instagram---the popular  photo-sharing social network---has a special section
of its site to acknowledge the importance of public code to the
company.\footnote{\url{https://www.instagram.com/about/legal/libraries}} In this page,
they thank the open source community for their contributions and explicitly list 25
open source libraries and frameworks used by the social
network.

Although open source has its origins in the eighties (or even earlier)~\cite{raymond1999cathedral}, the movement
is experiencing a renaissance period. One of the main reasons is the
appearance of modern platforms and workflows for developing and maintaining open source
projects~\cite{nadia2016roads}. The most famous example is GitHub; but other platforms are also relevant, such as Bitbucket
and GitLab. These platforms modernized the workflow
used on open source software development. Instead of changing e-mails with patches, developers contribute to a project by forking it, working and improving the code locally, and then submitting a pull
request to the project's leaders.

As a result, developers are creating open source code at
a rate never seen before. For example, today GitHub has more than \totalGitHubUsers\ users and
\totalGitHubRepositories\ repositories (without excluding forks). Consequently, these projects are
also {\em failing} at unprecedented rates.
Despite this fact, we have very few studies that investigate the {\em failures faced by open source
projects}~\cite{androutsellis2011open}.
We only find similar studies for commercial software.
For example, by means of a survey with developers and project managers,
Cerpa and Verner study the failure of 70
commercial projects~\cite{cerpa2009did}. They report that the most
common failures are due to unrealistic delivery dates, underestimated
project size, risks  not re-assessed through the project, and when staff is not
rewarded for working long hours. Certainly, these findings do not apply to
open source projects, which are developed without rigid schedules and
requirements, by groups of unpaid developers. The Standish Group's CHAOS
report is another study frequently mentioned by software
practitioners and  consultants~\cite{standish1994chaosreport}. The 2007 report mentions that 46\%
of software projects have cost and schedule problems and that 19\% are outright failures. Besides
having methodological problems, as pointed
by J\o{}rgensen and Mol\o{}kken-\O{}stvold~\cite{jorgensen2006large}, this report does not
target open source.

This paper describes an investigation with the maintainers of open source projects
that have failed, aiming to reveal the reasons for such failures, the maintenance practices
that distinguish failed projects from successful ones, the impact of failures on clients, and the strategies tried by maintainers to overcome the failure of
their projects. The paper addresses the following research questions:\\[-.25cm]

\noindent{\em RQ1: Why do open source projects fail?}
To answer this first RQ we select \totalStudiedProjectsWithoutCommits\ popular GitHub projects without any commits in the last year.
We complemented this selection with \totalStudiedProjectsWithDeprecatedMessages\ systems whose documentation explicitly mentions
that the project is abandoned. We asked the developers of these systems to describe the
reasons of the projects' failure. Finally, we categorize their responses into \totalReasonWhyOSSProjectFail\ major reasons.\\[-.25cm]

\noindent{\em RQ2: What is the importance of following a set of best open source maintenance practices?}
In this second research question, we check whether the failed projects used a set of best open source
maintenance practices, including practices to attract users and to
automate maintenance tasks, like continuous integration.\\[-.25cm]


\noindent{\em RQ3: What is the impact of the project failures?} To measure this impact, we counted the
number of  opened issues and pull requests of the failed projects and also the number of projects that
depend on them. The goal is to measure the impact of the studied failures, in terms of affected users, contributors,
and client projects.\\[-.25cm]

\noindent{\em RQ4: How do developers try to overcome the projects failure?}
In this last research question, we manually analyze the issues of the failed projects to collect
 strategies and procedures tried by their maintainers to avoid the failures.\\[-.25cm]

\begin{figure*}[!t]
  \centering
\vspace{-0.8mm}
\subfigure[ref1][Age]{\includegraphics[width=0.23\textwidth]{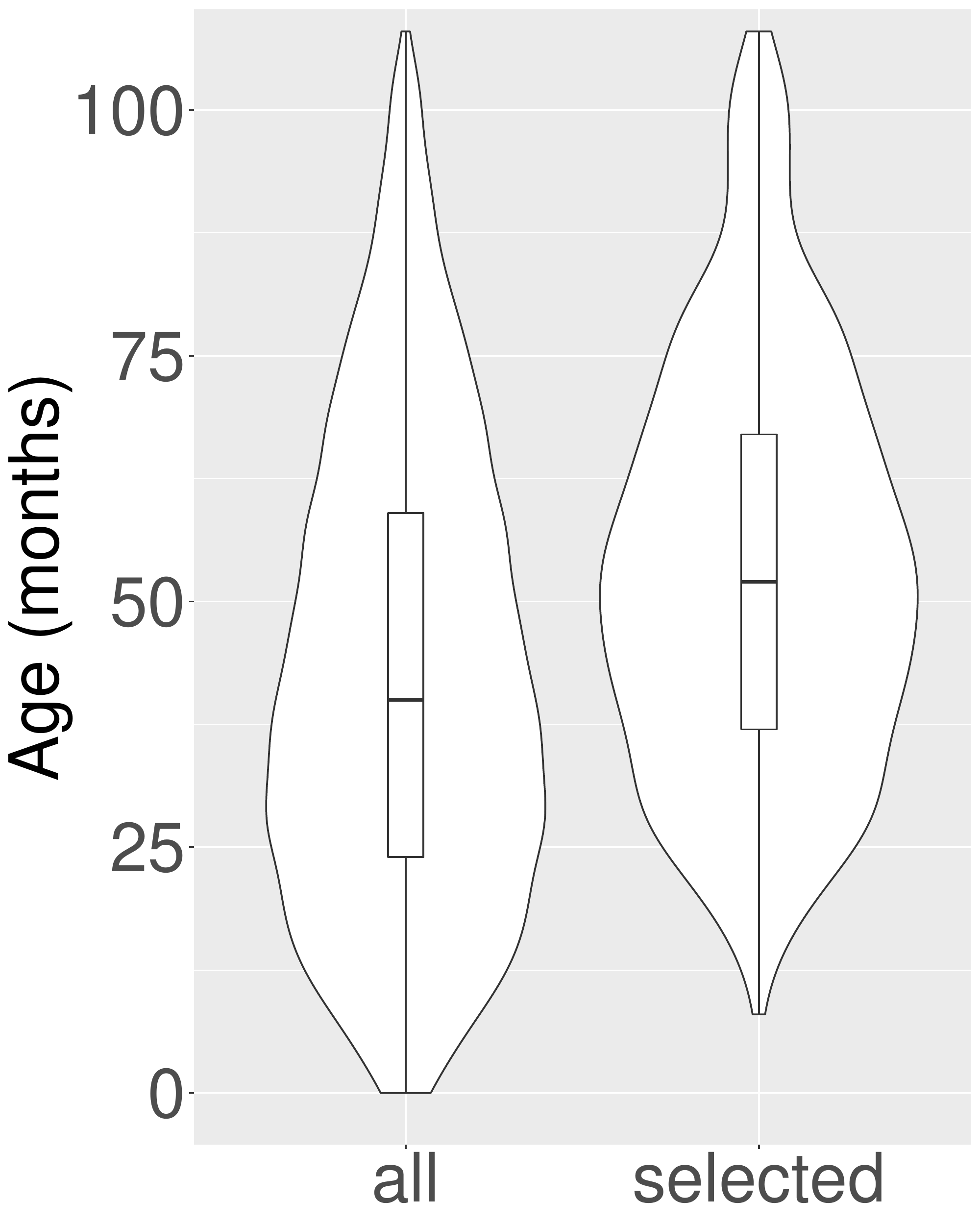}}
\quad
\subfigure[ref2][Contributors]{\includegraphics[width=0.23\textwidth]{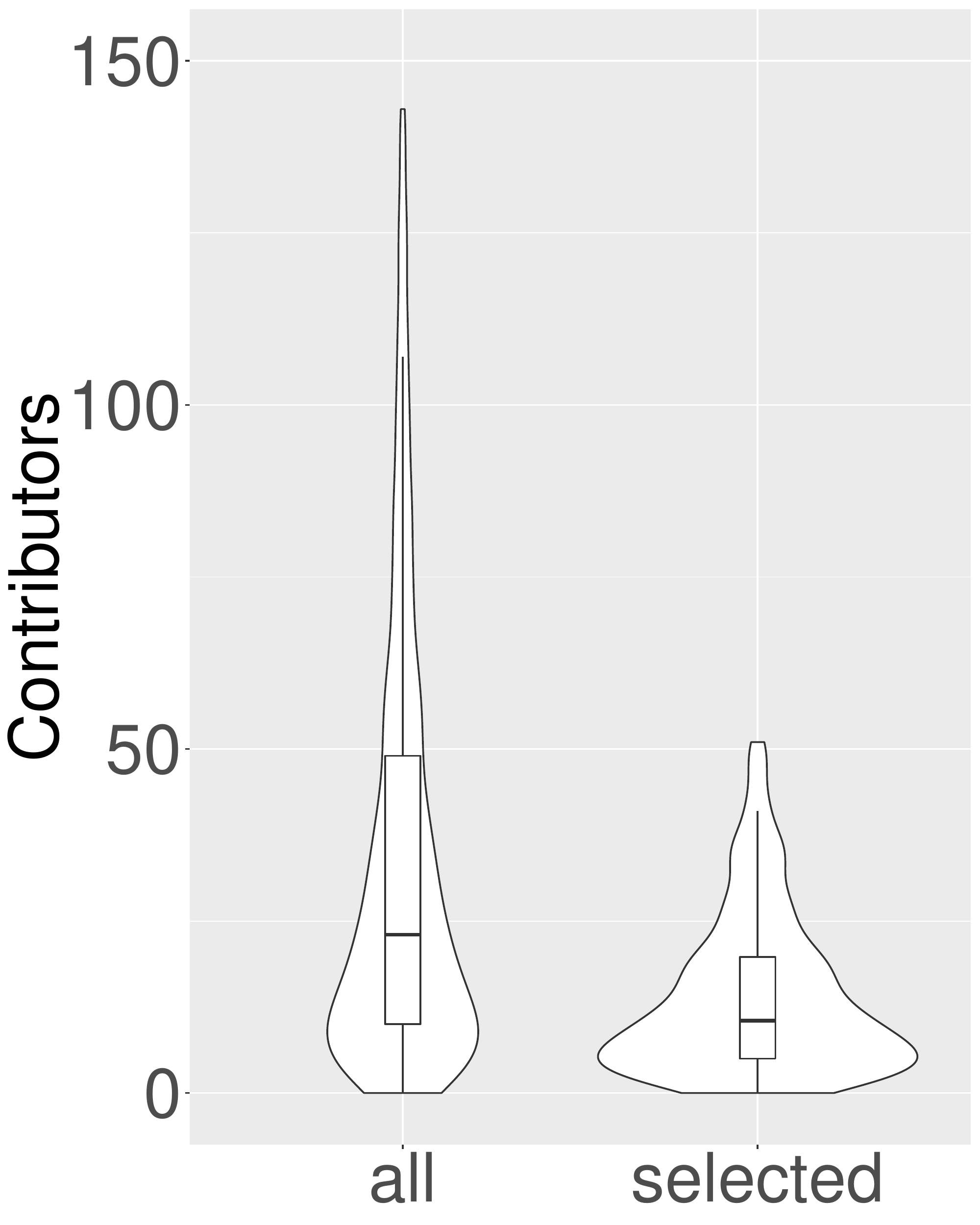}}
\quad
\subfigure[ref3][Commits]{\includegraphics[width=0.23\textwidth]{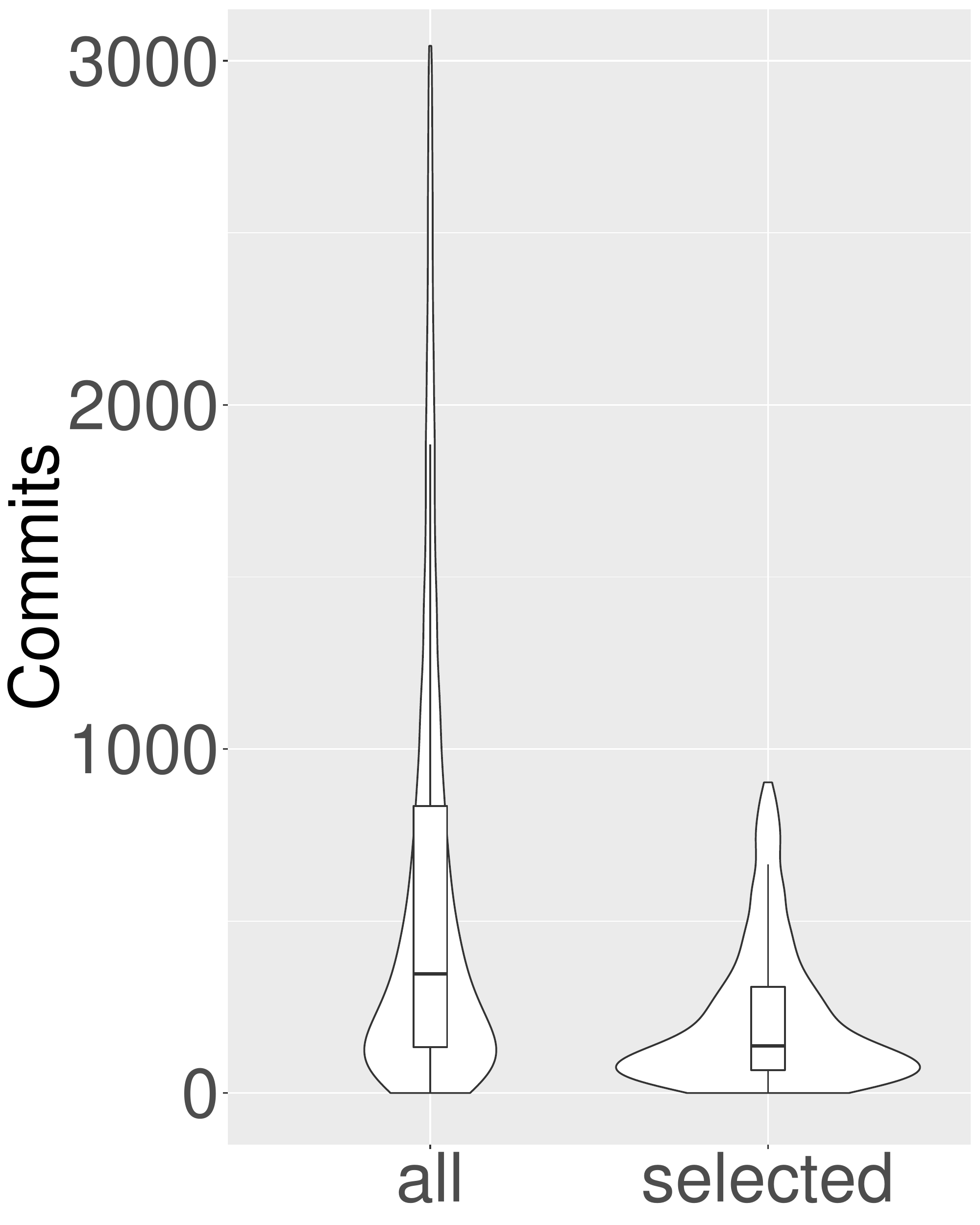}}
\quad
\subfigure[ref4][Stars]{\includegraphics[width=0.23\textwidth]{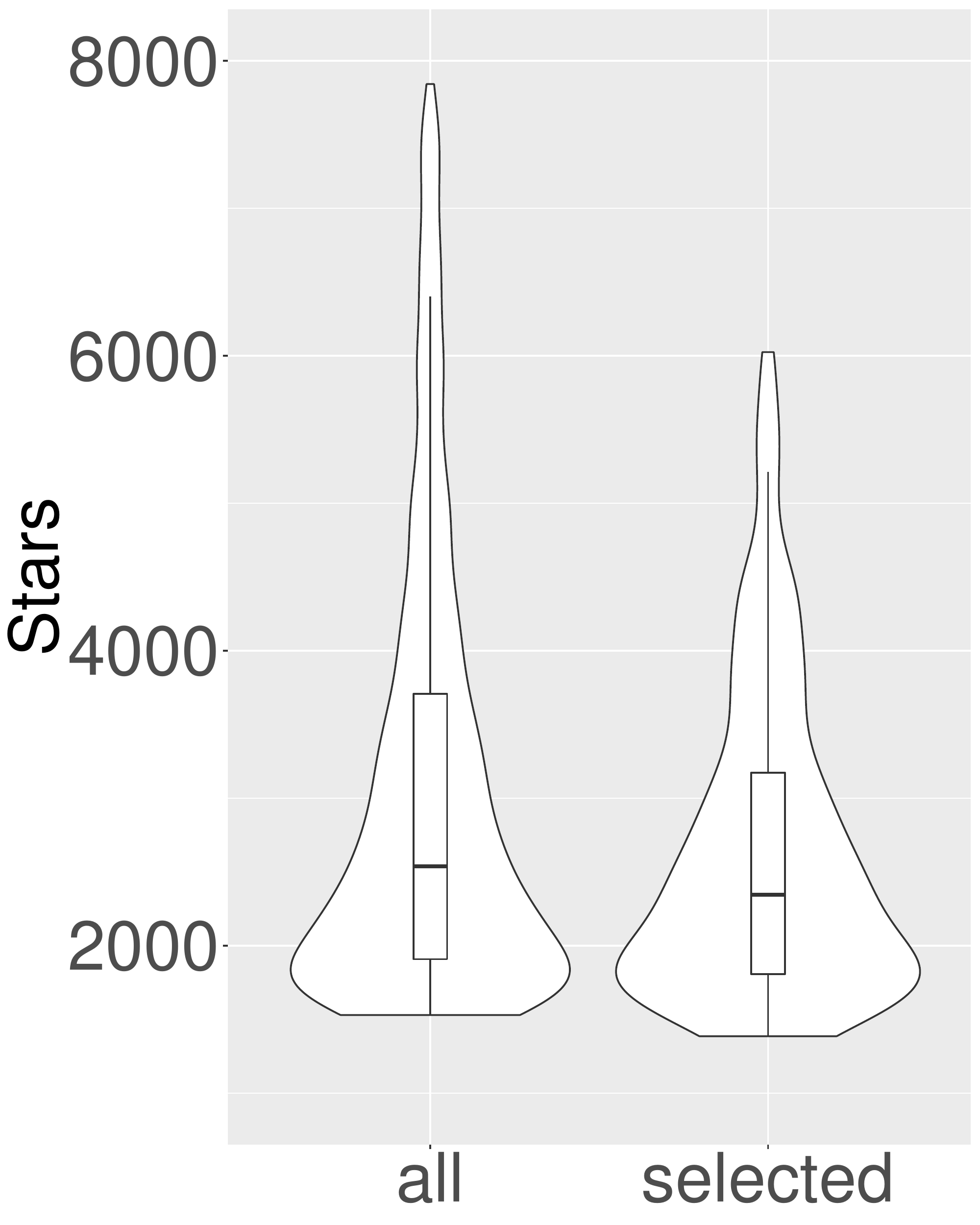}}
\vspace{-3mm}
\caption{Distribution of the (a) age, (b) contributors, (c) commits, and (d) stars, without outliers.}
\label{dataset_violinplot_projects_characteristics}
\end{figure*}

We make the following contributions in this paper:

\begin{itemize}
\item We provide a list of \totalReasonWhyOSSProjectFail\ reasons for failures in open source projects. By providing these reasons, using  data from real failures,
we intend to help developers to assess and control the risks faced by open source projects.\\[-.25cm]

\item We reinforce the importance of a set of best open source maintenance practices, by comparing their usage by the failed projects and also by the most and least popular systems in a sample of 5,000 GitHub projects.\\[-.25cm]

\item We document \issuesTotalStrategies\ strategies attempted by the maintainers of open source projects to overcome (without success) the failure of their projects.

\end{itemize}

We organize the remainder of the paper as follows. Section~\ref{sec:dataset} presents the dataset we use to
search for failed projects. Section~\ref{sec:why-projects-fail} to  Section~\ref{sec:overcome} presents answers
to each of the four research questions proposed in the study. Section~\ref{sec:discussion} discusses and puts our findings in a wider context.  Section~\ref{sec:threats} presents threats to validity; Section~\ref{sec:related-work} presents related work; and Section~\ref{sec:conclusion} concludes the paper.

\section{Dataset}
\label{sec:dataset}

The dataset used in this paper was created by first considering the top-5,000 most popular projects on GitHub (on
September, 2016). We use the number of stars as a proxy for popularity because it reveals how many people manifested
interest or appreciation to the project~\cite{borges2016icsme}. We limit the study to 5,000 repositories to focus on the maintenance challenges
faced by highly popular  projects.

We use two strategies to select systems that are no longer under maintenance
in this initial list of 5,000 projects. First, we select \totalProjectsWithoutCommitsBeforeFilter\ repositories (\totalProjectsWithoutCommitsBeforeFilterRate \%) without
commits in the last year. As examples, we have {\sc nvie/gitflow} (16,392~stars),
{\sc mozilla/BrowserQuest} (6,702~stars), and {\sc twitter/ty\-pe\-and.js} (3,750~stars).
Second, we search in the READ\-ME\footnote{
READ\-MEs are the first file a visitor is presented to when visiting a GitHub repository.
They include information on what the project does, why the project is useful, and
eventually the project status (if it is active or not).
} of the remaining repositories for terms
like~\aspas{deprecated},~\aspas{unmaintained}, ~\aspas{no longer maintained},~\aspas{no
longer supported}, and~\aspas{no longer under development}. We found such terms
in the READMEs of 207 projects (4\%). We then manually inspected these files to assure
that the messages indeed denote  inactive projects and to remove false
positives. After this inspection, we concluded that \totalStudiedProjectsWithDeprecatedMessages\ repositories (37\%) are true positives.
As an example, we have
{\sc google/gxui}\footnote{\url{https://github.com/google/gxui}}
whose README has this comment:\\[-.25cm]

\noindent{\em Unfortunately due to a shortage of hours in a day, GXUI is no longer maintained.}\\[-.25cm]

As an example of false positive, we have
{\sc twitter/labe\-lla.js}.\footnote{\url{https://github.com/twitter/labella.js}} In its README,
the following message initially led us to suspect that the project  is abandoned: \\[-.25cm]

\noindent{\em The API has changed. force.start() and \dots are deprecated.}\\[-.25cm]

However, in this case, deprecated refers to API elements and not to the project's status.
In a final cleaning step, we manually inspected the selected 704 repositores ($628 + 76$). We removed repositories that are not software projects
(\removedIsNotSoftwareProject\ repositories, e.g.,~books, tutorials, and awesome lists),  repositories whose native language is not English (\removedNativeLanguageIsNotEnglish\ repositories),
that were moved to another repository (\removedMovedToAnotherRepository\ repositories), 
and that are empty (\removedRepositoryEmpty\ repositories, which received their stars
 before being cleaned). We ended up with a list of \totalStudiedProjects\ projects 
 (542 projects without commits and 76 projects with an explicit deprecation message in the README).

Figure~\ref{dataset_violinplot_projects_characteristics} shows violin plots  with the distribution of age (in months), number of
contributors, number of commits, and number of stars  of the selected repositories. We provide plots for all 5,000
systems (labeled as {\em all}) and for the \totalStudiedProjects\ systems (12\%) considered in this study (labeled as {\em selected}).
The selected systems are older than the top-5,000 systems (\medianMonthsStudiedProjects\ vs \medianMonthsAllProjects\ months, median measures);
but they have less contributors (\medianContributorsStudiedProjects\ vs \medianContributorsAllProjects),
less commits (\medianCommitsStudiedProjects\ vs \medianCommitsAllProjects), and less stars (\medianStarsStudiedProjects\ vs \medianStarsAllProjects).
Indeed, the distributions are statistically different, according to the one-tailed variant of the Mann-Whitney U test (p-value $\leq$ 5\%). To show the effect size of this difference, we compute Cliff's delta
(or $d$). We found that the effect is small for age and commits, medium for contributors, and negligible for stars

GitHub repositories can be owned by a  person (e.g.,~\textsc{tor\-valds/li\-nux}) or by an organization
(e.g.,~\textsc{mozi\-lla/pdf.js}). In our dataset, \totalOrganizationProjects\ repositories (\totalOrganizationProjectsRate \%) are owed by organizations and \totalPersonalProjects\ repositories
(\totalPersonalProjectsRate \%) by users.  {JavaScript} is the most popular language (219 repositories, 36\%), followed by
Objec\-ti\-ve-C (98 repositories, 16\%), and {Java} (75 repositories, 12\%). In total, the dataset includes systems
spanning 26 programming languages. The first paper's author manually classified the application domain of the
systems in the dataset, as showed in Table~\ref{tab:domain-dataset}. There is a concentration on libraries and frameworks (\totalLibrariesAndFrameworkStudiedProject\ projects, \totalLibrariesAndFrameworkStudiedProjectRate \%),
which essentially reproduces a concentration also happening in the initial list of 5,000 projects.\footnote{For another research, we classified the domain of the top-5,000 GitHub projects;
59\% are libraries and frameworks.}\\[-.15cm]

\begin{table}[!ht]
\centering
\caption{Application domain of the selected projects}
\begin{tabular}{ l r l}
\toprule
{\bf Application Domain}    &  \multicolumn{2}{c}{\bf Projects}      \\
\midrule
Libraries and frameworks & \totalLibrariesAndFrameworkStudiedProject & \sbar{\totalLibrariesAndFrameworkStudiedProject}{\totalStudiedProjects} \\
Application software (e.g.,~text editors) & 63 & \sbar{63}{\totalStudiedProjects}\\
Software tools (e.g.,~compilers) & 31 & \sbar{31}{\totalStudiedProjects} \\
System software (e.g.,~databases) & 22 & \sbar{22}{\totalStudiedProjects} \\
\bottomrule
\end{tabular}
\label{tab:domain-dataset}
\end{table}

\noindent{\bf Dataset limitations:} The proposed dataset is restricted to popular open source
projects on GitHub. We acknowledge that there are popular
projects in other platforms, like Bitbucket, GitLab or that have their own version
control installations. Also, the  dataset does not include projects
that failed before attracting the attention of developers and users.
We consider less important to study such projects, since their failures 
did not have much impact. Instead, we focus on projects that succeeded to attract
attention, users, and contributors, but then failed, possibly impairing other
projects.

\section{Why do open source projects fail?}
\label{sec:why-projects-fail}

To answer the first research question, we conducted a survey with the developers of  \totalSentEmails\ open source projects with evidences of no longer being under maintenance.

\subsection{Survey Design}
\label{sec:survey-design}

The survey questionnaire has three open-ended questions: (1) Why did you stop maintaining the project? (2) Did you receive any funding to maintain the project? (3) Do you have plans to reactivate the project?
We avoid asking the developers directly about the reasons for
the project failures, because this question can lead to multiple
interpretations. For example, an abandoned
project could have been an outstanding learning experience to its
developers. Therefore, they might not consider that it has
failed. In Section~\ref{sec:combining-answers}, we detail the criteria
we followed to define that a project has failed based
on the answers to the survey questions.

Specifically to the developers of the \totalStudiedProjectsWithoutCommits\ repositories
without commits in the last year we added a first survey question, asking them to
confirm that the projects are no longer being maintained.
We also instructed them to only answer the remaining questions if they agree
with this fact. We sent the questionnaire to the repositories' owners or to the project's
principal contributor, in the case of repositories owned by organizations. Using this criterion,
we were able to find a public e-mail address
of \developersWithValidAndPublicEmail\ developers on GitHub.
However, \developersOwnerOrContributorOfTwoOrMoreProjects\ developers
are the owners---or the main contributors---of two or more projects. In this case,
we only sent one mail to these developers, referring to their first project in number
of stars, to avoid a perception of our mails as spam messages.

We sent the questionnaire to \totalSentEmails\ developers. After a period of 20 days, we obtained \developersSurveyedByEmail\ responses and \totalEmailsReturnedWithProblem\ mails returned due to the delivery problems,
resulting in a response rate of \responseRateSurvey\%, which is $\developersSurveyedByEmail / (\totalSentEmails - \totalEmailsReturnedWithProblem)$. To preserve the respondents'
anonymity, we use labels D1 to D\developersSurveyedByEmail\ to identify them. Furthermore, when quoting their answers we replace mentions to
repositories and owners by {\em [Project-Name]} and {\em [Project-Owner]}.
This is important because some answers  include critical comments
about developers or organizations.

Finally, for some projects, we found  answers to the first survey question (``Why did you stop maintaining the project?'') when inspecting their READMEs.
This happened with \developersSurveyedByReadme\ projects, identified by R1 to R\developersSurveyedByReadme. As an example, we have the following README:\\[-.25cm]

\noindent{\em Unfortunately, I haven't been able to find the time that I would like to dedicate to this project. (R6)} \\[-.25cm]

Therefore, for the first survey question, we collected \totalDevelopersSurveyed\ answers (\developersSurveyedByEmail\ answers by e-mail and \developersSurveyedByReadme\ answers from the projects' README).
We analyzed these answers using thematic
analysis~\cite{cruzes2011recommended,fse2016-why-we-refactor}, a
technique for identifying and recording ``the\-mes'' (i.e.,~patterns) in
textual documents. Thematic analysis  involves the following steps: (1)
initial reading of the answers, (2) generating
a first code for each answer, (3) searching for themes
among the proposed codes, (4) reviewing the themes to find opportunities
for merging, and (5) defining and naming the final themes. Steps (1) to (4)
were performed independently by each of the paper's authors. After this, a sequence
of meetings
was held to resolve conflicts and to assign the final themes (step 5).

\subsection{Survey Results}
\label{sec:survey-results}

This section presents the answers to the survey questions.
For the \developersSurveyedByEmail\ developers of systems with no commits in the last year,
the survey included an opening question asking if he/she agrees that the project
is no longer under maintenance. \developersAgreed\ developers (\developersAgreedRate \%) confirmed this project condition, as in the
following answer:\\[-.3cm]

\noindent{\em Yes, I surely have abandoned the project.} (D20)\\[-.3cm]


By contrast, \developersNotAgreed\ developers (\developersNotAgreedRate \%) did not agree with the project status.
For example, \developersKeepsMaintenanceOutOfRepository\ developers mentioned work being performed out of the main GitHub repository: \\[-.3cm]

\noindent{\em One current issue that does need to be resolved is that the entire site is served over https, but you wouldn't see that change in the repo.} (D18) \\[-.3cm]

\noindent{\em It is under maintenance. It's just not a lot of people are using it, and I am working on a new breaking version and thus didn't want to commit on the master branch.} (D30)\\[-.25cm]

Next, we present the reasons that emer\-ged after analysing the answers received for the first survey question (``Why did you stop maintaining the project?'').  We discuss each reason and give examples of answers associated to them.\\[-.3cm]

\noindent{\bf Lack of time:} According to \totalReasonOfLackOfTime\ developers,
they do not have free time to maintain the projects, as in the following
answers:\\[-.3cm]

\noindent{\em It was conceived during extended vacation. When I got back to working
I simply didn't have time. Building something like [Project-Name] requires
5-6 hours of work per day.} (D15)\\[-.3cm]

\noindent{\em I was the only maintainer and there was a lot of feature requests
and I didn't have enough time.} (D115)\\[-.25cm]

\noindent{\bf Lack of interest:} \totalReasonOfLackOfInterest\ developers answered they lost interest on the projects,
including when they started to work on other projects or domains, changed jobs, or were
fired.\footnote{Consequently, these developers do not have more time to work on their projects;
however, we reserve the lack of time theme to the cases where the developers still have
interest on the projects, but not the required time to maintain them.}
As examples, we have:\\[-.3cm]

\noindent{\em My interest began to wane; I moved to other projects.} (D67)\\[-.3cm]

\noindent{\em I'm not working in the CMS space at the moment.} (D77)\\[-.3cm]

\noindent{\em It became less professionally relevant/interesting.} (D80)\\[-.25cm]

\noindent{\em I was fired by the company that owns the project.} (D65)\\[-.25cm]

\noindent{\bf Project is completed:}  \totalReasonOfCompleted\ developers
consider that their projects are finished and do not need more features (just few and
sporadic bug fixes). As an example, we have the following answers:\\[-.3cm]

\noindent{\em Sometimes, you build something, and sometimes, it's done. Like if you built a
building, at some point in time it is finished, it achieved its goals. For [Project-Name] ---
it achieved all its goals, and it's done. \ldots The misconception is that people may mistake
an open source project with news. Sometimes there are just no more features to add,
no more news --- because the project is complete.} (D28)\\[-.3cm]

\noindent{\em I felt it was done. I think the
dominant idea is that you have to constantly update every open source project, but in
my opinion, this thing works great and needs no updates for any reason, and won't for
many, many years, since it's built on extremely stable APIs (namely git and Unix utilities).} (D69)\\[-.25cm]


\noindent{\bf Usurped by competitor:} \totalReasonOfUsurpedByCompetitor\ developers answered they abandoned the
project because a stronger competitor appeared in the market, as in the case of
these projects:\\[-.3cm]

\noindent{\em Google released ActionBarCompat whose goal was the same as
[Project-Name] but maintained by them.} (D2)\\[-.3cm]

\noindent{\em The project no longer makes sense. Apple has built technical and
legal alternatives which I believe are satisfactory.} (D71)\\[-.3cm]

\noindent{\em It's not been maintained for well over half a year and is formally
discontinued. There are better alternatives now, such as SearchView and
FloatingSearchView.} (R42)\\[-.25cm]

Specifically, 12 projects explicitly declare in their READMEs that they are no longer maintained due to the appearance of a strong competitor. In all cases, the update date of the project status as unmaintained occurred after appearing the competitor. For example, {\sc node-js-libs/node.io} was declared unmaintained four years after its competitor appeared. We also found this statement in its README: {\em I wrote node.io when node.js was still in its infancy.}\\[-.25cm]

\noindent{\bf Project is obsolete:} According to \totalReasonOfObsolete\ developers, the pro\-jects
are not useful anymore, i.e.,~their features are not more
required or applicable.\footnote{The theme does not include projects that are obsolete due to
outdated technologies, which have a specific theme.} As examples, we have the answers:\\[-.3cm]

\noindent{\em This was only meant as a stopgap to support older OSes. As we dropped that, we didn't need it anymore.} (D11)\\[-.3cm]

\noindent{\em I do not have an app myself anymore using that code.} (D36)\\[-.3cm]

\noindent{\em  I personally have no use for it in my work anymore.} (D38)\\[-.25cm]

\noindent{\bf Project is based on outdated technologies:} This reason, mentioned by \totalReasonOfOutdatedTehcnologies\ respondents, refer
to discontinuation due to outdated, deprecated or suboptimal technologies, including programming
languages, APIs, libraries, frameworks, etc. As examples, we have the following answers:\\[-.3cm]

\noindent{\em Due to Apple's abandonment of the Objective-C Garbage Collector which [Project-Name] relied heavily on, future development of [Project-Name] is on an indefinite hiatus.} (R20)\\[-.3cm]

\noindent{\em The core team is now building [Project-Name] in Dart instead of Ruby, and will no longer be maintaining the Ruby implementation unless a maintainer steps up to help.} (R34)\\[-.25cm]

\noindent{\bf Low maintainability:} This reason, as indicated by \totalReasonOfTechnicalReason\ developers, refers to maintainability problems. As examples, we have:\\[-.3cm]

\noindent{\em It is difficult to maintain a browser technology like JavaScript because browsers have very different quirks and implementations.} (D28)\\[-.3cm]

\noindent{\em The project reached an unmaintainable state due to architectural decisions made early in the project's life.} (D30)\\[-.25cm]

\noindent{\bf Conflicts among developers:} This reason, indicated by \totalReasonOfConflictsAmongDevelopers\ developers, denotes conflicts among developers or between developers and project owners, as in this answer:\\[-.3cm]

\noindent{\em The project was previously an official plugin---so the [Project-Owner] team worked with me to support it. However, they decided would not longer have the concept of plugins---and they ended the support on their side.} (D73)\\[-.25cm]

The remaining reasons include acquisition by a company, which created a private version of the project (\totalReasonOfAcquisition\ answers),
legal problems (\totalReasonOfLegalProblems\ answers), lack of expertise of the principal developer in the technologies used by the project (\totalReasonOfLackOfExpertise\ answer), 
and high demand of users, mostly in the form of trivial and meaningless issues (\totalReasonOfLackOfPatience\ answer).
Finally, in \totalReasonOfUnclear\ cases, it was not possible to infer a clear reason after reading the participant's answers. Thus, we classified these cases under an {\em unclear answer} theme. An example
is the following answer: {\em I am not so sure, but you can probably check the last commit details in GitHub.}

We also asked the participants a second question: {\em did you receive any funding to maintain the project?} \totalProjectsNotFunding\ out of \developersSurveyedByEmail\ answers (\totalProjectsNotFundingRate \%) were negative.
The positive answers mention funding from the company employing the respondent (\totalFundingCompanyEmploying\ answers), non-profit organizations (\totalFundingNonProfitOrganizations\ answers; e.g.,~{European Union}), and other private companies (\totalFundingPrivateCompany\ answers).
Finally, we asked a third question: {\em do you have plans to reactivate the project?}
Only \developersHavePlansToReactivateTheProject\ participants (\developersHavePlansToReactivateTheProjectRate \%) answered positively to this question.

\subsection{Combining the Survey Answers}
\label{sec:combining-answers}

In our study, we consider that a project has {\em failed} when at least one of the following conditions hold:\\[-.45cm]

\begin{enumerate}
\item The project is no longer under maintenance according to the surveyed developers and they do not have plans to reactivate the project (question \#3) and the project is not considered completed (question \#1).
\item The project documentation explicitly mentions that it is deprecated (without considering it completed).
\end{enumerate}

Among the considered answers, \totalProjectsFailuresReasonsAttendQuestionOne\ projects attend condition (1) and \totalProjectsFailuresReasonsAttendQuestionTwo\ projects attend condition (2).
The reasons for the failure of these projects are the ones presented in Section~\ref{sec:survey-results}, except when the themes are {\em lack of interest} or {\em lack of time}.
For these themes and when the answer comes from the top-developer of a project owned by an organization we made a final check on his number of commits.
We only accepted the reasons suggested by developers that are responsible for at least 50\% of the projects' commits.
For example, D85 answered he stopped maintaining his project due to a lack of time.
The project is owned by an organization and D85---although the top-maintainer of the project---is responsible for 30\% of the commits.
Therefore, in this case, we assumed that it would be possible to other developers to take over the tasks and issues handled by D85.
By applying this exclusion criterion, we removed \totalRemovedProjectsByHaveFewCommits\ projects from the list of projects.
The final list, which includes reasons for failures according to relevant top-developers or project owners, has \totalProjectsFails\ projects. In this paper, we call them {\em failed projects}.

Table~\ref{tab:why-stop} presents the reasons for the failure of these projects.
The most common reasons are project was {\em usurped by competitor} (\totalReasonWhyFailOfUsurpedByCompetitor\ projects), project is {\em obsolete} (\totalReasonWhyFailOfObsolete\ projects), {\em lack of time} of the main contributor (\totalReasonWhyFailOfLackOfTime\ projects), {\em lack of interest } of the main contributor (\totalReasonWhyFailOfLackOfInterest\ projects), and project is based on {\em outdated technologies} (\totalReasonWhyFailOfOutdatedTehcnologies\ projects).
It is also important to note that projects can fail due to multiple reasons, which happened in the case of \totalProjectsFailWithMultipleReasons\ projects.
Thus, the sum of the projects in Table~\ref{tab:why-stop} is \totalReasonFailsSum\
(and not \totalProjectsFails\ projects).\footnote{The values in Table~\ref{tab:why-stop} are not exactly the ones
presented in Section~\ref{sec:survey-results} due to the inclusion and exclusion criteria defined in this section.}

\begin{table}[!h]
\centering
\caption{Why open source projects fail?}
\begin{tabular}{ l lc l }
\toprule
{\bf Reasons}	&  {\bf Group} & \multicolumn{2}{c}{\bf Projects}     \\
\midrule
Usurped by competitor	& Environment & 	\totalReasonWhyFailOfUsurpedByCompetitor	& \sbar{\totalReasonWhyFailOfUsurpedByCompetitor}{\totalReasonFailsSum}	 \\
Obsolete	 & 	Project & \totalReasonWhyFailOfObsolete		& \sbar{\totalReasonWhyFailOfObsolete}{\totalReasonFailsSum}	 \\
Lack of time	 & 	Team & \totalReasonWhyFailOfLackOfTime	& \sbar{\totalReasonWhyFailOfLackOfTime}{\totalReasonFailsSum}	 \\
Lack of interest	 & 	Team & \totalReasonWhyFailOfLackOfInterest	& \sbar{\totalReasonWhyFailOfLackOfInterest}{\totalReasonFailsSum}	 \\
Outdated technologies	 & 	Project & \totalReasonWhyFailOfOutdatedTehcnologies		& \sbar{\totalReasonWhyFailOfOutdatedTehcnologies}{\totalReasonFailsSum}	 \\
Low maintainability	 & 	Project & \totalReasonWhyFailOfTechnicalReason		& \sbar{\totalReasonWhyFailOfTechnicalReason}{\totalReasonFailsSum}	 \\
Conflicts among developers	 & Team & 	\totalReasonWhyFailOfConflictsAmongDevelopers	 & \sbar{\totalReasonWhyFailOfConflictsAmongDevelopers}{\totalReasonFailsSum}	 \\
Legal problems	 & Environment & 	\totalReasonWhyFailOfLegalProblems	 & \sbar{\totalReasonWhyFailOfLegalProblems}{\totalReasonFailsSum}	 \\
Acquisition	 & Environment & 	\totalReasonWhyFailOfAcquisition	 & \sbar{\totalReasonWhyFailOfAcquisition}{\totalReasonFailsSum}	 \\

\bottomrule
\end{tabular}

\label{tab:why-stop}
\end{table}

As presented in Table~\ref{tab:why-stop}, we classified the reasons for failures in three groups:
(1) reasons related to the development team (including lack of time,
lack of interest, and conflicts among developers); (2) reasons related to
project characteristics (including
project is obsolete, project is based on outdated technologies,
and low project maintainability);
(3) reasons related to the environment where
the project and the development team are placed (including
usurpation by competition, acquisition by a company, and legal issues).\\[-.25cm]

\begin{tcolorbox}[left=0mm,right=0mm,boxrule=0.25mm,colback=gray!5!white]
{\em Summary:} Modern open source projects fail due to reasons related to project characteristics (\totalReasonProjectCharacteristics\ projects; e.g.,~low maintainability), followed
by reasons related to the project team (\totalReasonDevelopmentTeam\ projects; e.g.,~lack of time or
interest of the main contributor); and due to environment reasons (\totalReasonExternalEnvironment\ projects; e.g.,~project was usurped by a competidor or legal issues).
\end{tcolorbox}

\newcommand\xbar[2]{#1  {\color{darkgray} \rule{\dimexpr #2pt * 18}{5.5pt}}{\color{lightgray} \rule{\dimexpr 16pt - (#2pt * 18)}{5.5pt}}}

\begin{table*}[h]
\centering
\caption{Percentage of projects following practices recommended when maintaining GitHub repositories. The effect size reflects the extent of the difference between the repositories in a given group (Top, Bottom, or Random) and the failed projects}
\label{fails-characteristics}
\begin{tabular}{ l | r | rl | rl | rl }
\toprule
\textbf{Maintaince Practice}			& \multicolumn{1}{c|}{\textbf{Failed}}		& \multicolumn{1}{c}{\textbf{Top}}	& {\bf Effect}	&  \multicolumn{1}{c}{\textbf{Bottom}}	& {\bf Effect}	& \multicolumn{1}{c}{\textbf{Random}}	& {\bf Effect}	\\
\midrule
README                   	& \xbar{99}{0.99}				& \xbar{100}{1}		& \multicolumn{1}{c|}{-}	& \xbar{100}{1}		& \multicolumn{1}{c|}{-}	& \xbar{100}{1}		&  \multicolumn{1}{c}{-}				\\
License                  	& \xbar{61}{0.61}       			& \xbar{88}{0.88}	& small				& \xbar{60}{0.60}	& \multicolumn{1}{c|}{-}	& \xbar{73}{0.73}	& \multicolumn{1}{c}{-}					\\
Home Page                 	& \xbar{58}{0.58}       			& \xbar{87}{0.87}	& small 			& \xbar{52}{0.52}	& \multicolumn{1}{c|}{-}	& \xbar{60}{0.60}	& \multicolumn{1}{c}{-}					\\
{Continuous Integration}     	& \xbar{27}{0.27}       			& \xbar{68}{0.68}	& medium			& \xbar{41}{0.41}	& \multicolumn{1}{c|}{-}	& \xbar{45}{0.45}	& small							\\
Contributing                	& \xbar{16}{0.16}       			& \xbar{72}{0.72}	& large				& \xbar{13}{0.13}	& \multicolumn{1}{c|}{-}	& \xbar{32}{0.32}	& small							\\
Issue Template              	& \xbar{0}{0}           			& \xbar{15}{0.15}	& small				& \xbar{2}{0.02}	& \multicolumn{1}{c|}{-}	& \xbar{5}{0.05}	& \multicolumn{1}{c}{-}					\\
Code of Conduct                	& \xbar{0}{0}           			& \xbar{13}{0.13}	& \multicolumn{1}{c|}{-}	&\xbar{0}{0}		& \multicolumn{1}{c|}{-}	& \xbar{2}{0.02}	& \multicolumn{1}{c}{-}					\\
Pull Request Template          	& \xbar{0}{0}           			& \xbar{3}{0.03} 	& \multicolumn{1}{c|}{-}	&\xbar{0}{0}		& \multicolumn{1}{c|}{-}	& \xbar{0}{0}		& \multicolumn{1}{c}{-}					\\
\bottomrule
\end{tabular}
\end{table*}


\section{What is the importance of open source maintenance practices?}
\label{sec:characteristics-of-fails}

In this second question, we investigate whether the failed projects followed (or not) a set of best open source maintenance practices, which are recommended when hosting projects on GitHub.\footnote{\url{https://opensource.guide}}
Section~\ref{sec:rq2-methodology} describes the methodology we followed to answer the research question and Section~\ref{sec:rq2-results} presents the results and findings.

\subsection{Methodology}
\label{sec:rq2-methodology}

We analyzed four groups of projects: the \totalProjectsFails\ projects that have failed, as described in Section~\ref{sec:combining-answers} ({\em Failed}), the top-\totalProjectsFails\ and the bottom-\totalProjectsFails\ projects by number of stars ({\em Top} and {\em Bottom}, respectively), and a random sample of \totalProjectsFails\ projects ({\em Random}). {\em Top}, {\em Bottom}, and {\em Random} are selected from the initial sample of top-5,000 projects, described in Section~\ref{sec:dataset}, and after applying the same cleaning steps defined in this section. The rationale is to compare the {\em Failed} projects with the most popular projects in our dataset, which presumably should follow most practices; and also with the least popular projects and with a random sample of projects.

For each project in the aforementioned groups of projects we collected the following information:\footnote{Five of these maintenance practices are explicitly recommended at: \url{
https://help.github.com/articles/helping-people-contribute-to-your-project}}
 (1) presence of a README file (which is the landing page of GitHub repositories); (2) presence of a separate file with the project's license; (3) availability of a dedicated site and URL to promote the project, including examples, documentation, list of principal users, etc; (4) use of a continuous integration service (we check whether the projects use Travis CI, which is the most popular CI service on GitHub, used by more than 90\% of the projects that enable CI, according to a recent study~\cite{hilton2016usage}); (5) presence of a specific file with guidelines for repository
contributors; (6) presence of an issue template (to instruct developers to write issues according to the repository's guidelines); (7) presence of a specific file with a code of conduct (which is a document that establishes expectations for the behavior of the project's participants~\cite{tourani2017code}); and (8) presence of  a pull request template (which is a document to instruct developers to submit pull requests according to the repository's guidelines).

After collecting the data for each project in each group we compared
the obtained distributions. First, we analyzed the statistical
significance of the difference between the {\em Top}, {\em Bottom},
and {\em Random} groups vs the {\em Failed} group, by applying
the Mann-Whitney test at p-value = 0.05. To show the effect
size of the difference, we used Cliff's delta. Following the guidelines of previous work~\cite{grissom2005effect, tian2015mobile, linares2013api}, we interpreted the
effect size as small for $0.147 < d < 0.33$, medium for
$0.33 \leq d < 0.474$, and large for $d \geq 0.474$.

\subsection{Results}
\label{sec:rq2-results}

Table~\ref{fails-characteristics} shows the percentage of projects following each practice. 
Despite the group, the most followed practices are the presence of a README file, the presence of a license file, and the availability of a project home page. 
For example, for the {\em Failed} projects the percentage of projects following these practices are 99\%, 61\%, and 58\%, respectively. For the {\em Top} projects, the same values are 100\%, 88\%, and 87\%, respectively. 
The least followed practices are issue templates, code of conduct, and pull request templates. 
We did not find a single project following these  practices in the {\em Failed} group. 
By contrast, 15\%, 13\%, and 3\% of the {\em Top} projects have these three kind of documents, respectively. 
In general, we observe the following order among the groups of projects regarding the adoption of the eight considered practices: $\mathit{Top} > \mathit{Random} > \mathit{Failed} \equiv \mathit{Bottom}$. 
In other words, there is a relevant adoption of most practices by the {\em Top} projects. 
By contrast, the \totalProjectsFails\ projects that failed and that are studied in this paper are more similar to the {\em Bottom} projects. 
This fact is reinforced by the analysis of Cliff's delta coefficient. 
There is a {\em large} effect size between the adoption of contributing guidelines by the {\em Top} (72\%) and the {\em Failed} projects (16\%), and a {\em medium} difference in the case of continuous integration services (68\% vs 27\%). 
For licenses, home pages, and issue templates, the difference is {\em small}. For the remaining practices, the difference is negligible or does not exist in statistical terms. 
In the case of the {\em Bottom} projects, there is no statistical difference for the eight considered documents.
Finally, for  {\em Random}, there is a {\em small} difference when we consider the use of continuous integration and contributing guidelines.

\begin{tcolorbox}[left=0mm,right=0mm,boxrule=0.25mm,colback=gray!5!white]
{\em Summary:}
Regarding the adoption of best open source maintenance practices, the failed projects are more similar to the least popular projects  than to the most popular ones. Therefore, these practices seem to have an effect on the success or failure of open source projects. The practices with the most relevant effects are contributing guidelines (large), continuous integration (medium), and licences, home pages, and issue templates (small).
\end{tcolorbox}

\section{What is the Impact of Failures?}
\label{sec:impact}

With this third research question, we intend to assess the impact of the failure of the studied projects, both to end-users and to the developers of client systems.
First, we present the approach we used to answer the question (Section~\ref{sec:impact-approach}). Then, we present the results (Section~\ref{sec:impact-results}).

\subsection{Methodology}
\label{sec:impact-approach}

To answer the question, we collected data on (a) the number of issues and pull requests of the failed projects; and (b) the number of systems that depend on these projects, according to data provided by GitHub and by a popular JavaScript package manager.\\[-.25cm]

\begin{figure*}[!t]
  \centering
\vspace{-.8mm}
\subfigure[ref1][Opened issues]{\includegraphics[width=\lenghtReasons \textwidth]{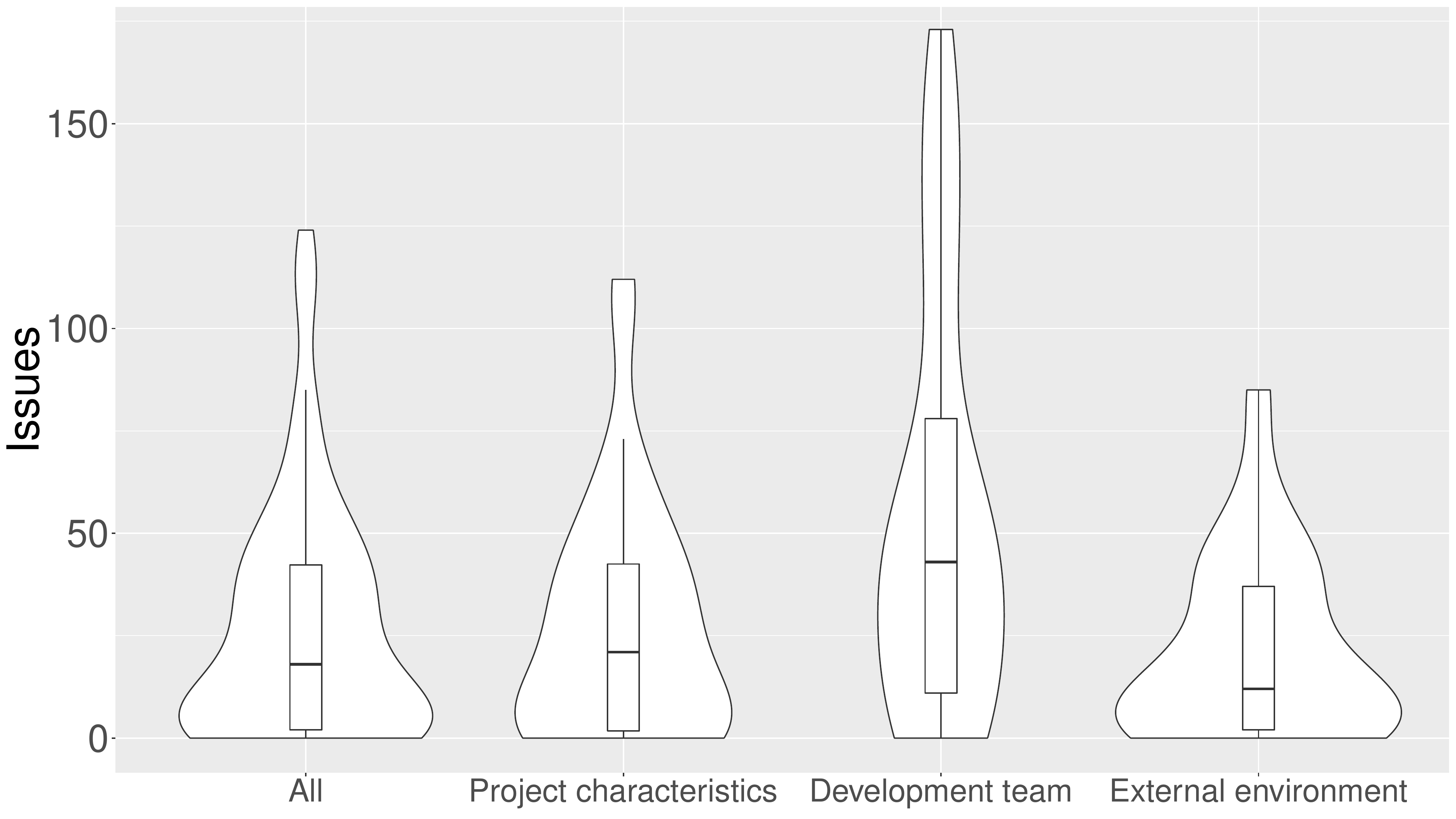}}
\quad
\subfigure[ref2][Opened pull requests]{\includegraphics[width=\lenghtReasons \textwidth]{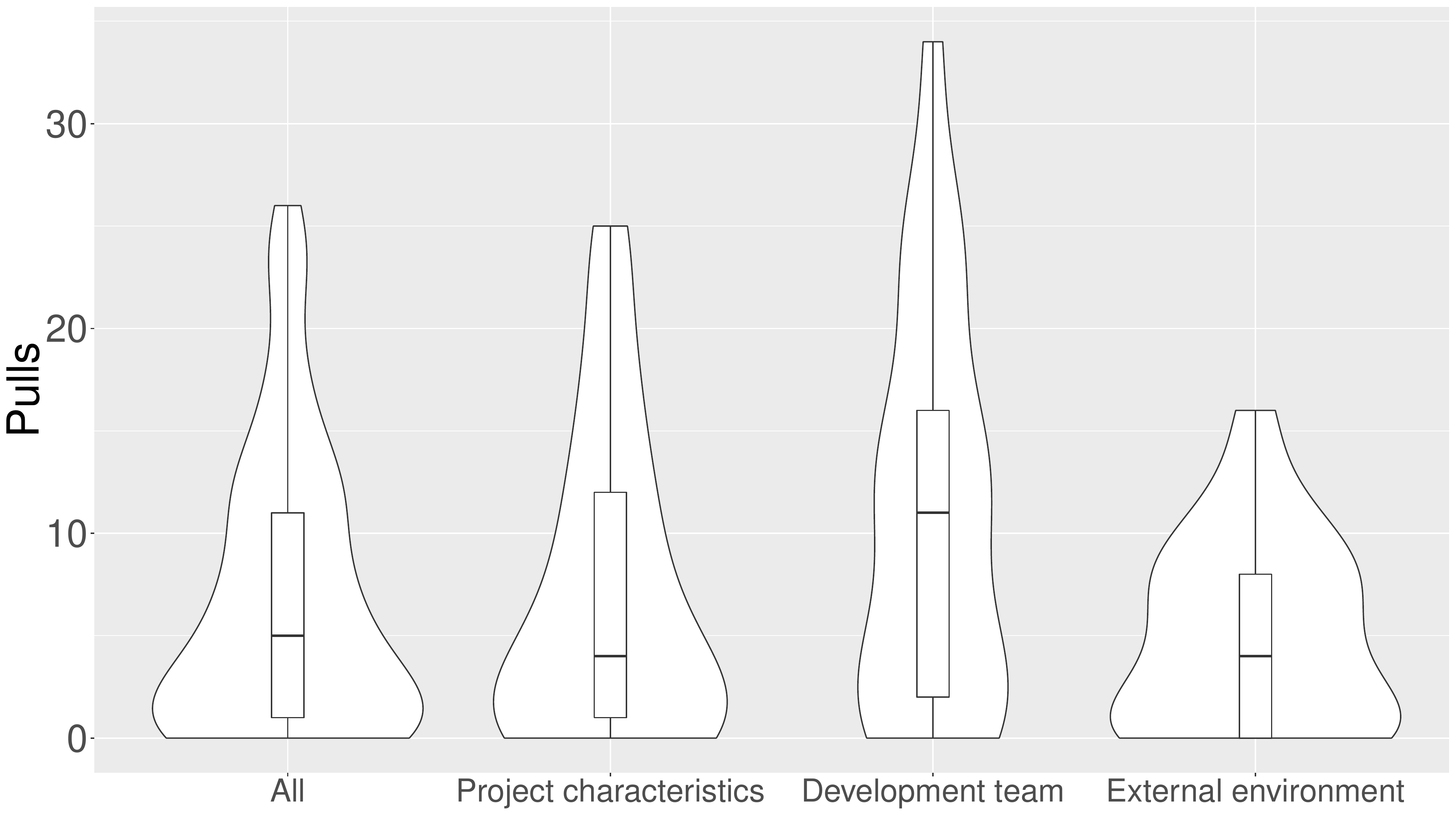}}
\vspace{-3mm}
\caption{Distribution of the (a) Opened issues and (b) Opened pull requests, without outliers}
\label{violinplot_issues_pulls_opened}
\end{figure*}
\subsection{Results}
\label{sec:impact-results}

\noindent{\em Issues and Pull Requests:} Using the GitHub API, we collected the number of opened issues and opened pull requests for each failed project (in the case of issues, we excluded \issuesTotalProjectsWichNotManageByGitHub\ projects that do not use GitHub to handle issues).
Our rationale is that one of the negative impacts of an abandoned project is a list of bugs and enhancements (issues) that will not be considered and a list of source code modifications (pull requests) that will not be implemented.
Pending issues impact the projects' users, who need to keep using a project with bugs or a
frozen set of features, or who will have to migrate to other projects.
Pending pull requests contribute to the frustration of the projects' contributors, who will not have their effort appreciated.\\[-.25cm]

\noindent{\em Dependencies:} We also collected data on
projects that depend on the failed projects and that therefore
are using unmaintained systems. To collect dependencies data, we first rely on a
GitHub service that reports the number of client repositories that depend on
a given repository.\footnote{\url{https://github.com/blog/2300-visualize-your-project-s-community}}
Unfortunately, this feature is available only to Ruby systems.
To cover more projects, we also consider dependency data provided
by {\sc npm}, a popular package manager
for JavaScript. As result, we analysed the dependencies of \totalJavaScriptRubyProjectsFails\  projects,
including  \totalRubyProjectsFails\ Ruby projects and \totalJavaScriptProjectsFails\ JavaScript ones.
\\[-.25cm]

\noindent{\em Issues and Pull Requests:}  Figure~\ref{violinplot_issues_pulls_opened} shows violin plots with the distributions of opened issues and pull requests. 
Considering the \issuesTotalFailProjectsWichManageIssuesByGitHub\ failed projects with issues on GitHub, the median number of opened issues is \medianIssuesOpenedAll\ and the median number
 of pending pull requests is \medianPullsOpenedAll.
The top-3 failed projects with the highest number of pending issues have \topOneIssuesOpened, \topTwoIssuesOpened, and \topThreeIssuesOpened\ issues.
The top-3 failed projects with the highest number of pending pull requests have \topOnePullsOpened, \topTwoPullsOpened, and \topThreePullsOpened\ pull requests.
The figure also shows the number of opened issues and pull requests grouped by failure reasons.
The median number of issues for failures associated to project characteristics, development team, and environment reasons are \medianIssuesOpenedProjectCharacteristics, \medianIssuesOpenedDevelopmentTeam, and \medianIssuesOpenedExternalEnvironment\ issues, respectively.
For pull requests, the median measures for the same groups of reasons are \medianPullsOpenedProjectCharacteristics, \medianPullsOpenedDevelopmentTeam, and \medianPullsOpenedExternalEnvironment\ pull requests, respectively.
By applying Kruskal-Wallis test to compare multiple samples, we find that these distributions are not different. \\[-.25cm]

\noindent{\em Dependencies:} \totalRubyProjectsFailsWithoutDependents\ (out of \totalRubyProjectsFails)
Ruby repositories do not have dependent projects. However, we also found projects
with \totalDependentsTopOneRubyProjectsFails, \totalDependentsTopTwoRubyProjectsFails,  \totalDependentsTopThreeRubyProjectsFails\, and \totalDependentsTopFourRubyProjectsFails\ dependents. Regarding the JavaScript systems,
\totalJavaScriptProjectsFailsOutNPM\ (out of \totalJavaScriptProjectsFails) projects do not have an entry on {\sc npm} (although {\sc npm} is
 very popular, systems can use other package managers or do not use a
 package manager at all). \totalJavaScriptProjectsWithFiveOrLessDependents\ projects have five or less dependents and
 three systems have respectively \totalDependentsTopOneJavaScriptProjectsFails, \totalDependentsTopTwoJavaScriptProjectsFails, and \totalDependentsTopThreeJavaScriptProjectsFails\ dependents.\\[-.25cm]

\begin{tcolorbox}[left=0mm,right=0mm,boxrule=0.25mm,colback=gray!5!white]
{\em Summary:} The failed projects have \medianIssuesOpenedAll\ opened issues and \medianPullsOpenedAll\  opened pull requests (median measures).
\totalJavaScriptProjectsWithFiveOrLessDependentsRate \% of the Ruby and JavaScript projects have less than five
dependents, which suggests that most clients have also abandoned
these projects.
\end{tcolorbox}

\section{How do developers try to overco\-me the projects failure?}
\label{sec:overcome}

In this fourth research question, we qualitatively investigate attempts
to overcome the failure of the studied projects.


\subsection{Methodology}
\label{sec:rq4-methodology}

The first paper's author read the 20 most recent opened issues
and the 20 most recent closed issues of each of the \totalProjectsFails\ failed projects
(in a total of 1,654 issues).
As a result, he collected \issuesTotalIssuesAboutProjectStatus\ issues where the developers
question the status of the projects and/or
discuss alternatives to restart the development. The issues,
which are identified by I1 to I\issuesTotalIssuesAboutProjectStatus, cover \issuesTotalIssuesAboutProjectStatus\ projects in the list of
 failed projects. Examples
of titles of selected issues are: {\em Is this project dead?}
(I18), {\em Is this project maintained?} (I1), and {\em Is
development of this ongoing?} (I7). After this first step,
the first author extracted a set of recurrent strategies (or ``themes'') suggested by
developers to overcome the failure of the projects the issues refer to.
The proposed themes were  validated by the second paper's author,
in a last step.

\subsection{Results}
\label{sec:rq4-results}

After analyzing the issues, we found \issuesTotalStrategies\ strategies tried by owners or collaborators to
overcome the unmaintained status of the projects. Next, we describe these strategies.\\[-.25cm]

\noindent{\bf Moving to an organization account:}  This strategy, mentioned in \issuesSuggestsOrganization\ issues,
refers to the creation of an organization account with a name similar to the project's name.
The hope is that with this kind of account it would be easier to attract new maintainers
and to manage permissions. As examples, we have these comments:\\[-.25cm]

\noindent{\em Would creating a [Project-Name] repo in a [Project-Name] org be something people would want?} (I31)\\[-.25cm]

\noindent{\em I am totally cool with setting up an org and transferring
control... Just let me know what you need.} (I3)\\[-.15cm]

\noindent{\bf Transfer the project to new maintainers:} This strategy, discussed for
\issuesSuggestsNewMaintainer\ projects, consists in a complete transfer of the project's maintainership
to other developers (but keeping the project's name), as discussed in these issues:\\[-.25cm]

\noindent{\em Who want to take over this project will be appreciated.
We will watch the project together for a while and I will grant every permission.} (I10) \\[-.25cm]

In  \issuesSuggestsNewMaintainerAndFound\ projects, a new developer was found and assumed
the project, as documented in this issue:\\[-.25cm]


\noindent{\em I have started working on {\em [Project-Name]}.
{\em [Project-Owner]} transferred the repository to my account.} (I7) \\[-.15cm]

We tracked the activity of the new maintainers, until February, 2017. They did not
perform significant contributions to the projects, despite minor commits. In one project, we found the following complaint about
the new maintainer:\\[-.25cm]

\noindent{\em @[Owner-Name] gave this repo to someone who has never been active on GitHub, so this repo is basically dead.} (I11)\\[-.25cm]

\noindent{\bf Accepting new core developers:} In \issuesSuggestsAddCollaborators\ cases, to overcome the low
activity on the repositories, volunteers offered to help with the maintenance,
as core developers. For example, we have this issue:\\[-.25cm]


\noindent{\em @[Project-Name] Would you be open to adding more collaborators to this repo?} (I17)\\[-.25cm]

In all cases, the proposals were not
answered or accepted. As an example, we have this project owner, who requested
a detailed maintenance plan before accepting the maintainer: \\[-.15cm]

\noindent{\em I'd be willing to do this if the collaborators
provided a roadmap of what they'd like to accomplish with the library.} (I17)\\[-.15cm]

Although it is not exactly an overcome strategy, in \issuesSuggestsSubstitute\ cases
owners suggested the developers to start collaborating on another project,
as in this issue:\\[-.25cm]

\noindent{\em I'd suggest you look at [Project-Name]. It's very active and modern. I'm trying to find time to switch over myself.} (I9)\\[-.25cm]


Finally, although the presented strategies were not able to restart the development of the studied projects, they should not be considered as completely failed ones. 
To illustrate this fact, we selected 348 projects that almost failed in the year before the study (they have five or less commits).
182 projects (52\%) indeed failed in the next year (the studied one). 
However, 35 projects show evidences of recovering (they have more than the first quartile of commits/year in the studied year, i.e.,~15 commits). 
After inspecting the documentation and issues of these 35 projects, we found that 14 projects attracted new core developers (third strategy), two were transferred to new maintainers (second strategy), and two projects moved to an organization account (first strategy). 

\begin{tcolorbox}[left=0mm,right=0mm,boxrule=0.25mm,colback=gray!5!white]
{\em Summary:} Developers attempted  \issuesTotalStrategies\ strategies to overcome the failure of
their projects: (a) moving to an organization account; (b) transfer the project to new maintainers;
(c) accepting new core developers.
\end{tcolorbox}

\subsection{Complementary Investigation: Forks}
\label{sec:forks}

Forks are used on GitHub to create copies of repositories. The mechanism allows developers to make changes to a
project (e.g., fixing bugs or implementing new features) and submit the modified code back to the
original repository, by means of pull requests. Alternatively, forks can become independent
projects, with their own community of developers. Therefore, forks can be used to
overcome the failure of projects, by bootstrapping a new project from the
codebase of an abandoned one. For this reason, we decided to complement the investigation
of RQ4 with an analysis of the forks of the failed projects.

\begin{figure}[!h]
  \centering
\subfigure[ref1][Forks]{\includegraphics[width=\lenghtForks \textwidth]{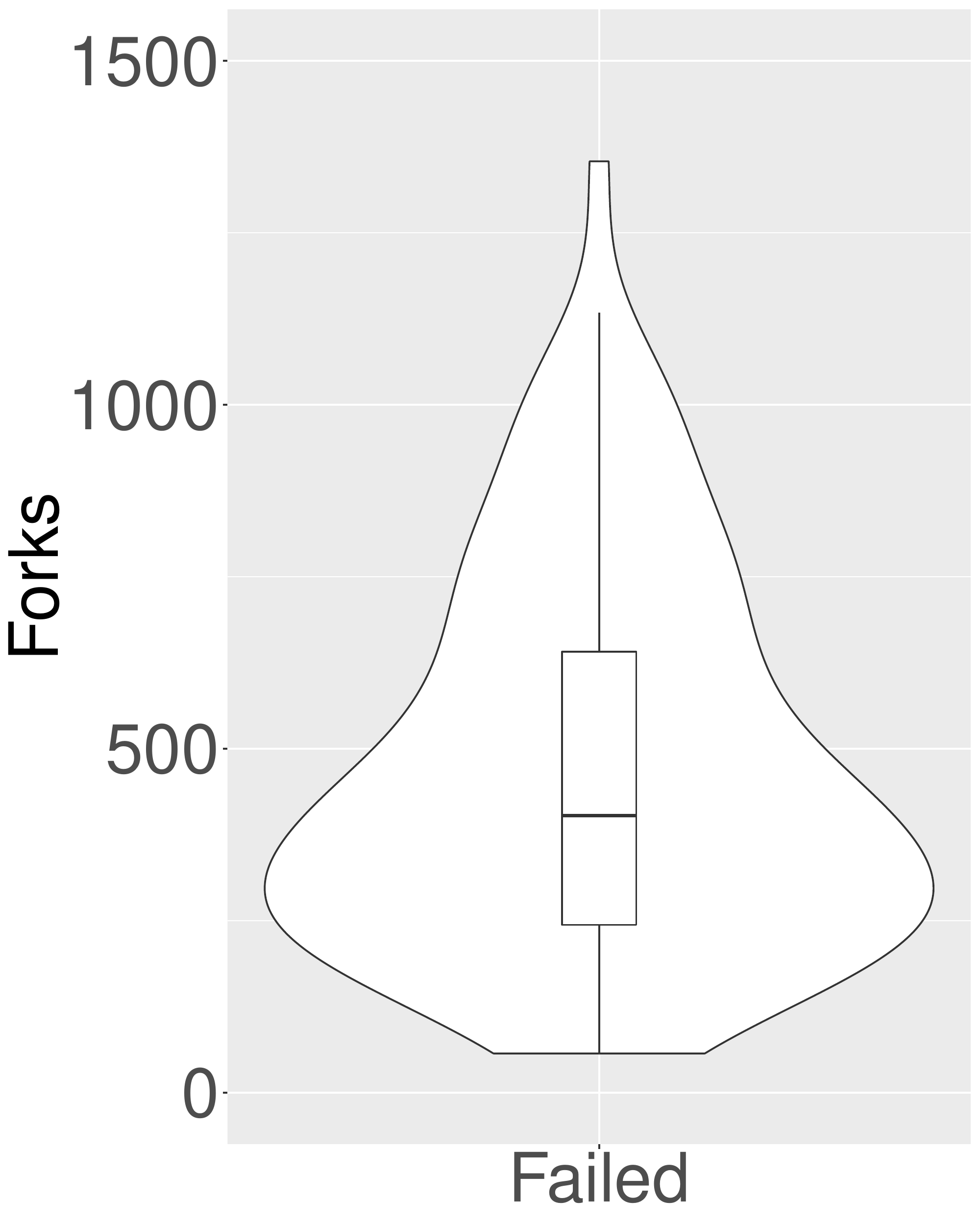}}
\quad
\subfigure[ref2][Stars (best fork)]{\includegraphics[width=\lenghtForks \textwidth]{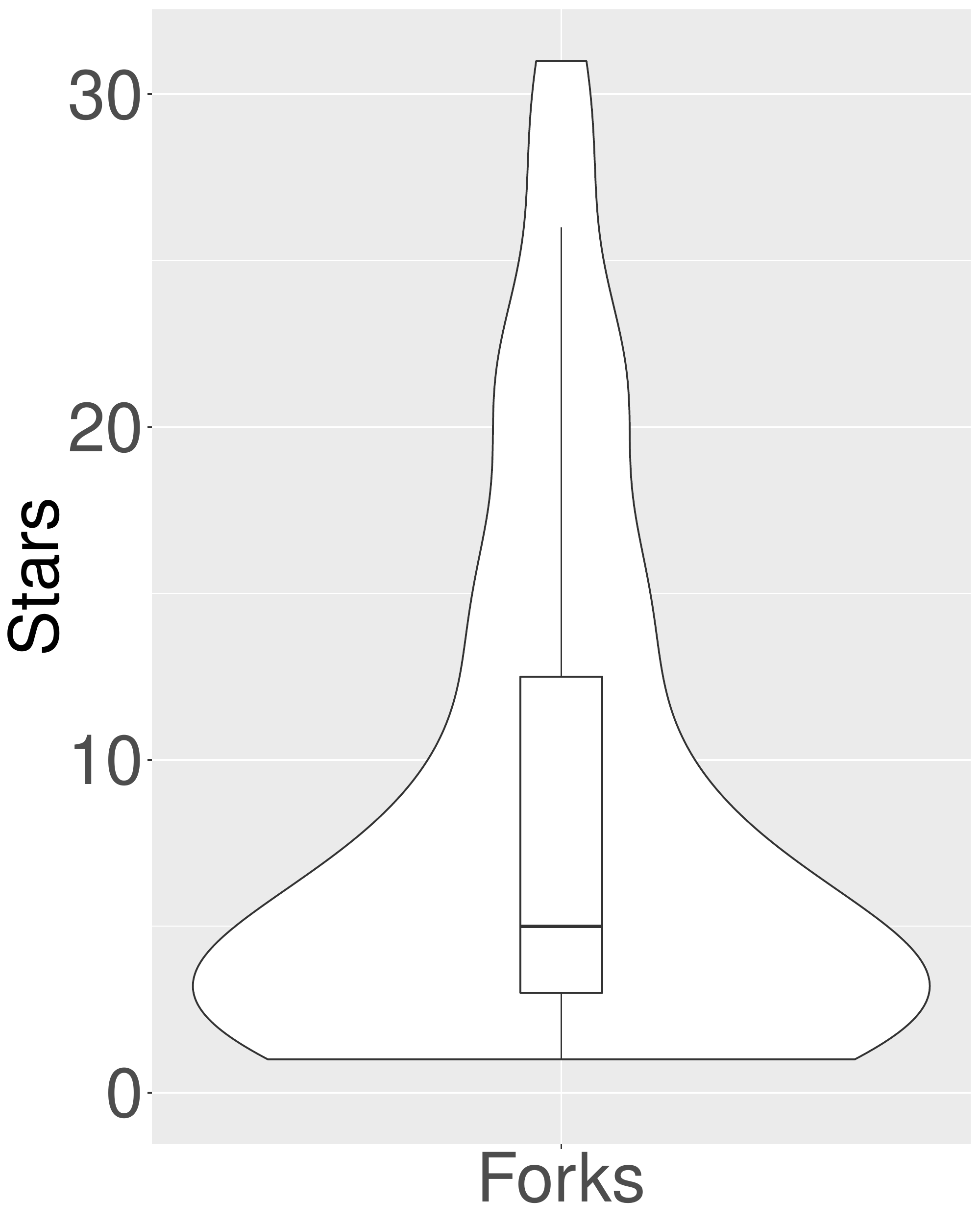}}
\vspace{-3mm}
\caption{Distribution of the (a) number of forks of the failed projects and (b) number of stars of the fork with the highest number of stars, for each failed project; both violin plots without outliers}
\label{fig:forks}
\end{figure}

Figure~\ref{fig:forks}a shows the distribution of the number of forks of the failed projects. They usually have
a relevant number of forks, since it is very simple to fork projects on GitHub. The first, median, and third quartile
measures are \forksFirstQuartile, \forksMedian, and \forksThirdQuartile\ forks, respectively. The violin plot in Figure~\ref{fig:forks}b aims to reveal the relevance of these forks.
For each project, we computed the fork with the highest number of stars. The violin plot shows the
 distribution of the number of stars of these most successful forks. As we can see, most forks are not
 popular at all. They are probably only used to submit pull requests or to create a copy of a repository, for backup
 purposes~\cite{jiang2016and}. For example, the third quartile measure is \forksThirdQuartileOfStarOfBestFork\ stars. However, there are two systems
 with an outlier behavior. The first one is an audio player, whose fork has 1,080 stars. In our survey, the
 developer of the original project answered that he abandoned the project due to other interests. However, his
 code was used to fork a new project, whose README acknowledges that this version {\em ``is a substantial rewrite of the fantastic work done in version 1.0 by [Projet-Owner] and others''}. Besides 1,080 stars, the forked project has 70 contributors
 (as February, 2017). The second outlier is a dependency injector for Android, whose
 fork was made by Google and has 6,178 stars. The forked project's README mentions that it ``is currently in active development, primarily internally at Google, with regular pushes to the open source community".\\[-.1cm]
 

\begin{tcolorbox}[left=0mm,right=0mm,boxrule=0.25mm,colback=gray!5!white]
{\em Summary:} Forks are rarely used on GitHub
to continue the development of open source projects that have failed.
\end{tcolorbox}

\section{Discussion}
\label{sec:discussion}

In this section, we discuss the main findings of our study.\\[-.15cm]

\noindent{\bf Completed projects and first Law of Software Evolution:} An interesting finding of the
survey with developers is the category of completed projects (\totalReasonOfCompleted\  systems, \totalReasonOfCompletedRate \%), which
are considered feature-completed by their developers. They do not plan
to evolve these systems, because ``{\em adding more features would just
obfuscate the original intent}'' (D55) of the projects. Moreover,
they also think the projects will not need adaptive maintenance,
as in this answer:\\[-.25cm]

\noindent{\em I just stopped working on it because what I have works very well, and will continue working very well until Unix stops being the foundation of most Web development, which basically means until the end of the human race \ldots Most projects don't build on similarly solid foundations so they probably need to change more often.} (D56)\\[-.25cm]

Someone can argue that these projects contradict the first Law
of Software Evolution, which prescribes that ``programs are never completed''~\cite{lehman1980programs}.
However, Lehman's Laws only
apply to E-type systems, where the
``E'' stands for evolutionary.\footnote{The first law (Continuing Changing) is as follows: ``An E-type system must be continually adapted, else it becomes progressively
less satisfactory in use.''~\cite{lehman1997metrics}} In these systems,
the environment around the program changes and hence the requirements and the program specification~\cite{herraiz2013evolution}.
Therefore, Lehman opens the possibility to have completed programs, when they target an
environment controlled by the developers or that is very stable
 (e.g.,~the Unix ecosystem, as mentioned by Developer D56).\\[-.25cm]

\noindent{\bf Competition in open source markets:}
The study reveals an important competition between open source projects. The most common reason for project failures is
the appearance of a stronger open source competitor (\totalReasonWhyFailOfUsurpedByCompetitor\ projects).
Usually, this competitor is the major organization
responsible for the ecosystem the project is inserted on, specifically Google
(Android ecosystem, \totalUsurpedByGoogle\ projects) and Apple (iOS ecosystem,
\totalUsurpedByApple\ projects). Therefore, open source developers should be aware of the risks
of starting a project that may attract the attention of major players,
particularly when the projects have a tight integration and dependency with established
platforms, like Android and iOS. Clients should also evaluate the risks of using
these ``non-official'' projects. They should evaluate if it is worth to accept the opportunity
costs of delaying the use of a system until it is provided as a built-in service.
Alternatively, they can conclude that the costs of delaying the adoption further exceeds
the additional benefits of providing earlier a service to end-users.
Other competitors mentioned in the survey are {\sc d3/d3} (a visualization library
for JavaScript) and MVC frameworks, also for JavaScript, such as {\sc facebook/react}.
For example, one developer mentioned that ``{\em high-end front-end development
seems to be moving away from jQuery plugins''} (D18). This result
confirms that web development is a competitive domain, where the
risks of failures are considerable, even for highly popular projects.\\[-.25cm]

\noindent{\bf Practical implications:} This study provides insights to the definition of lightweight
``maturity models'' to open source projects. By lightweight, we mean that
such models should be less complex and detailed than equivalent models for commercial
software projects, like CMMI~\cite{chrissis2003cmmi}. But at least they can prescribe that
open source projects should manage and constantly assess the risk factors that emerged from
our empirical investigation. We shed light on three particular factors: (a) risks associated to development teams
(for example, projects than depend on a small number of core developers may fail due to the lack of
time or lack of interest of these developers, after a time working in the project); (b) risks
associated to the environment the projects are immersed (which seems to be particularly
relevant in the case of projects with a tight integration with mobile operating systems or in the case
of web libraries and frameworks); (c) risks associated to project characteristics and decisions, like
the use of outdated technologies. Furthermore, we also showed the importance of practices
normally recommended to open source development on GitHub. We 
show that successful projects provide documents like README,
contributing guidelines, usage license declarations, and issue templates. They also
include a separate home page, to promote the projects among end-users. Finally,
we showed evidences on the benefits provided by continuous integration,
in terms of automation of tasks like compilation, building, and testing.

\section{Threats To Validity}
\label{sec:threats}

The threats to validity of this work are as follows:\\[-.25cm]

\noindent{\bf External Validity:}  Threats to external validity were partially discussed when
presenting the dataset limitations (Section~\ref{sec:dataset}). We complement this discussion as follows.
First, when investigating the use of continuous integration by the failed, top, bottom, and random projects
(RQ2,  Section~\ref{sec:characteristics-of-fails}), we only consider the use of Travis CI. However, Travis is the most popular CI service
on GitHub, used by more than 90\% of the repositories that enable CI~\cite{hilton2016usage}.
Second, the investigation of dependent projects (RQ3, Section~\ref{sec:impact})
only considered systems implemented in Ruby and JavaScript. For JavaScript, we only
analyzed dependency data provided by a single package manager system ({\sc npm}).
\\[-.25cm]

\noindent{\bf Internal Validity:} The first threat relates to the selection of the survey participants.
We surveyed the project owner, in the case of repositories owned by individuals,
or the developer with the highest number of commits, in the case
of repositories owned by organizations. Although experts on their projects,
it is possible that some participants omitted in their answers the
real reasons for the project failures. To mitigate this threat, we avoid asking the participants
directly about the causes of the project failures.
A second threat relates to the themes denoting reasons for
project failures (RQ1) and strategies on how to overcome them (RQ4). We acknowledge
that the choice of these themes is to some extent subjective. For example, it is possible that
different researchers reach a different set of reasons, than the ones proposed in Section~\ref{sec:survey-results}.
To mitigate this threat, the initial selection of themes in RQ1 was
performed independently by the two authors of this paper. After this initial proposal, daily
meetings were performed during a whole week to refine and improve the initial selection.
In the case of RQ4, the themes were proposed by the first paper's author and validated
by the second author.
A third internal validity threat might appear when interpreting the results of RQ2. In this
case, it is important to consider that association does not imply in causation. For example,
by just providing contributing guidelines or codes of conduct, a project does not necessarily will succeed.\\[-.25cm]

\noindent{\bf Construct Validity:}  A first construct validity threat relates to thresholds and parameters
used to define the survey sample. We consider as unmaintained
the projects that did not have a single commit in the last year (Section~\ref{sec:dataset}). We recognize a threat in the selection of this threshold and time frame.
However, to mitigate this threat, we included in the survey \developersSurveyedByReadme\
projects whose README explicitly declares that the project is unmaintained or deprecated.
The second threat concerns the data about the maintenance practices used
to answer RQ2 (Section~\ref{sec:characteristics-of-fails}). This data was collected automatically, by means of scripts that rely
on regular expressions to match different names and extensions used by the
documents of interest (e.g.,~license.md and license.txt). However, we cannot guarantee that the implemented
expressions match all possible variations of file names. Moreover, we did not investigate and check
the quality of the retrieved documents. For example, we consider that a project has contributing
guidelines when this document exists in the repository and it is not empty.

\vspace{-.1cm}

\section{Related Work}
\label{sec:related-work}

Capiluppi et al.~\cite{capiluppi2003characteristics} analyze 406 projects from
FreshMeat (a deprecated open source repository).
For each project, they compute a set of measures
along four main dimensions: community of developers,
community of users, modularity and documentation,
and software evolution. They report that most
projects (57\%) have one or two developers and that
only a few (15\%) can be considered active,
i.e.,~continuing improving their popularity and
number of users and developers. However, they do not
investigate the reasons for the project failures.
Khondhu et al.~\cite{khondhu2013all} discuss the attributes and
characteristics of inactive projects on SourceForge.
They report that more than 10,000 projects are
inactive (as November, 2012). They also compare the
maintainability of inactive projects with
other project categories (active and dormant), using
the maintainability index (MI)~\cite{oman1992metrics}. They conclude
that the majority of inactive systems
are abandoned with  a similar or increased
maintainability, in comparison to their initial
status. However, there are serious
concerns on using MI as a maintainability predictor~\cite{bijlsma2012faster}.

Tourani et al.~\cite{tourani2017code} investigate the
role, scope and influence of codes of conduct
in open source projects. They report that
seven codes are used by most projects,
usually aiming to provide a safe and inclusive
community, as well as dealing with diversity issues.
After surveying the literature on empirical studies
aiming to validate Lehman's Laws, Fernandez-Ramil
et al.~\cite{Fernandez-Ramil2008} report that most works conclude
that the first law (Continuing Change)
applies to mature open source projects. However, in this work
we found \totalReasonOfCompleted\ completed projects, according to their developers. These
projects deal with stable requirements and environments and
therefore do not need constant updates or modifications.

Ye and Kishida~\cite{ye2003toward} describe a study to understand what motivates developers
to engage in open source development. Using as case study the
GIMP project (GNU Image Manipulation Program) they  argue that learning
is the major driving force that motivates people to get involved in
open source projects. However, we do not known if this find applies to the
new generation of open source systems, developed using platforms
as GitHub. Eghbal~\cite{nadia2016roads} reports on the risks and challenges
to maintain modern open source projects.
She argues that open source plays a key role
in the digital infrastructure that sustain our
society today. But unlike physical
infrastructure, like bridges and roads,
open source still lacks a
reliable and sustainable source of funding. Avelino et al. 
concluded that nearly two-thirds of a sample of 133 popular GitHub projects depend on one or two developers to survive~\cite{icpc2016}.

Humphrey~\cite{humphrey2005big} presents 12 reasons for project failures, but in
the context of commercial software  and to justify
the adoption of maturity models, like CMMI~\cite{chrissis2003cmmi}.
The reasons are presented and explained in the form of questions concerning
why large software projects are hard to manage, the kinds of management
systems needed, and the actions required to implement such systems.
Lavallee et al.~\cite{lavallee2015good} weekly observed during ten months
the development of software projects in a large telecomunnication company.
They show that organization factors, e.g., structure and culture, have a major impact
on the success or failure of software projects. However, in our study these factors
did not appear with the same importance. For example, only three projects failed due to conflicts among developers. We hypothesise  this is
due to the decentralized and  community-centric characteristics of open
source code. Washburn et al.~\cite{washburn2016went} analyse 155
postmortems published on the gaming site Gamasutra.com.
They report the best practices and
common challenges faced in game development and provide a list
of factors that impact project outcomes. For example, they
found that the creativity of the development team is
often a relevant factor in the success or failure of a game.
As a practical recommendation, they mention that projects
should practice good risk management techniques.
We argue that the failure factors elicited in this paper
are a start point to include such practices in open source
development, i.e.,~to control the risks we need first to
known them.

Recent research on open source
has focused on the organization of
successful open source proj\-ects~\cite{mockus2002two}, on how to attract and
retain newcomers~\cite{zhou2015will, steinmacher2016overcoming, leeunderstanding, pinto2016more, canfora2012going},
and on specific features provided by GitHub, such as pull
requests~\cite{gousios2014exploratory, gousios2015work, gousios2016work}, forks~\cite{jiang2016and}, and stars~\cite{borges2016icsme, borges2016promise}.

\vspace{-.15cm}

\section{Conclusion}
\label{sec:conclusion}

In this paper, we showed that
the top-5 most common
reasons for the failure of open source projects are: project was usurped by competitor (\totalReasonWhyFailOfUsurpedByCompetitor\ projects), project
became functionally obsolete (\totalReasonWhyFailOfObsolete\ projects), lack of time of the main contributor (\totalReasonWhyFailOfLackOfTime\ projects),
lack of interest of the main contributor (\totalReasonWhyFailOfLackOfInterest\ projects), and project based on outdated technologies
(\totalReasonWhyFailOfOutdatedTehcnologies\ projects). We also showed that there is an important difference between the failed projects
and the most popular and active projects on GitHub, in terms of following best open source
maintenance practices. This difference is more important regarding the
availability of contribution guidelines and the use of continuous integration. Furthermore, the failed projects
have a non-negligible number of opened issues and pull requests.
Finally, we described three
strategies attempted by  maintainers to overcome
the failure of their projects.

As future work,
we propose that researchers and practitioners work on defining and validating ``maturity
models'' for open source projects, which can contribute  to minimize the risks of adopting these
projects in practice. We also recommend investigation on proactive strategies to avoid the
failure of projects, for example by identifying and recommending new maintainers with the
required expertise to work in projects under threats of being deprecated.

\section*{Acknowledgments}

We would like to thank the \developersSurveyedByEmail\ GitHub developers who took
their time to answer our survey.
This research is supported by grants from FAPEMIG, CAPES, and CNPq.

\balance

\bibliographystyle{ACM-Reference-Format}

\bibliography{fse2017}

\end{document}